\documentclass[12pt]{article}
 
\newcommand{\mrm}[1]{\mathrm{#1}}
\newcommand{\pT}{p_{\perp}}
\newcommand{\pTmin}{p_{\perp\mrm{min}}}
\newcommand{\alphaem}{\alpha_{\mrm{em}}}
\newcommand{\gtrsim}{\raisebox{-0.8mm}{$\stackrel{>}{\sim}$}}

\renewcommand{\b}{\mathrm{b}}
\renewcommand{\c}{\mathrm{c}}
\renewcommand{\d}{\mathrm{d}}
\newcommand{\e}{\mathrm{e}}
\newcommand{\g}{\mathrm{g}}
\newcommand{\p}{\mathrm{p}}
\newcommand{\q}{\mathrm{q}}
\newcommand{\s}{\mathrm{s}}
\renewcommand{\u}{\mathrm{u}}
\newcommand{\pbar}{\overline{\mathrm{p}}}
\newcommand{\qbar}{\overline{\mathrm{q}}}
\newcommand{\Jpsi}{\mrm{J}/\psi}

\newenvironment{Itemize}{\begin{list}{$\bullet$}%
{\setlength{\topsep}{0.2mm}\setlength{\partopsep}{0.2mm}%
\setlength{\itemsep}{0.2mm}\setlength{\parsep}{0.2mm}}}%
{\end{list}}
\newcounter{enumct}
\newenvironment{Enumerate}{\begin{list}{\arabic{enumct}.}%
{\usecounter{enumct}\setlength{\topsep}{0.2mm}%
\setlength{\partopsep}{0.2mm}\setlength{\itemsep}{0.2mm}%
\setlength{\parsep}{0.2mm}}}{\end{list}}

\newlength{\abstwidth}
\begin{document}
 
\sloppy

\begin{flushright}
LU TP 95--35 \\
December 1995
\end{flushright}
 
\vspace*{1cm}
 
\begin{center}
{\LARGE\bf Soft Photoproduction Physics
\footnote{to appear in the proceedings of the Durham Workshop on
HERA Physics, ``Proton, Photon and Pomeron Structure'', 
17--23 September 1995, Durham, U.K.}}\\[10mm]
{\large Torbj\"orn Sj\"ostrand} \\[2mm]
{\em Department of Theoretical Physics,} \\[1mm]
{\em University of Lund, S\"olvegatan 14A,} \\[1mm]
{\em S-223 62 LUND, Sweden}\\[15mm]
\end{center}

\begin{center}
ABSTRACT \\[2mm]
\begin{minipage}{\abstwidth}
Several topics of interest in soft photoproduction physics are
discussed. These include jet universality issues (particle flavour
composition), the subdivision into event classes, the buildup of the
total photoproduction cross section and the effects of multiple 
interactions.  
\end{minipage}
\end{center}

\section{Introduction}

Why should we be concerned with soft photoproduction? The most
obvious answer is that it is there, at an overwhelmingly large
cross section, and so it is an irresistible challenge to 
unravel what is going on. The nature of a real photon is not 
well understood --- its multifaceted behaviour is illustrated by
the conventional but handwaving subdivision into direct and 
resolved components. Furthermore, the bremsstrahlung 
spectrum off an electron beam contains photons of varying 
virtuality, which prompts a study of the poorly explored 
transition region between a real photon and the virtual photon 
of DIS.

The ``problems'' of photoproduction should not be seen in isolation.
The relation between total and diffractive cross sections in
$\gamma$p events is part of the same pomeron physics complex
that is now under intense scrutiny in the context of rapidity gaps 
in DIS; the study of the transition region between real and virtual 
photons, in particular, may well provide further clues to the gap 
production mechanism. The total $\gamma$p cross section is also 
related to the issue of multiple interactions and from there to the
small-$x$ behaviour of parton distributions and the influence of
hot spots. The similarities between the photon and hadrons will allow
various cross-checks between pp/p$\pbar$, $\gamma$p and $\gamma\gamma$
physics. Jet universality issues can be explored by comparing 
high-$\pT$ (hard) with low-$\pT$ (soft) jets.

Finally, as a general philosophical comment, today perturbative QCD
is well established. Further calculations certainly are very useful
for precision physics, but may not bring many basic new insights.
By contrast, the nonperturbative aspects still contain many 
challenges and opportunities. 

A breakdown of the physics of photoproduction events might look 
something like:
\begin{Itemize}
\item The total cross section of $\gamma$p events is subdivided into a 
direct and a resolved component. 
\item The resolved class is induced by a spectrum of virtual fluctuations
$\gamma \leftrightarrow \q\qbar$, where the low-virtuality part
can be associated with low-mass vector mesons such as $\rho^0$,
$\omega$ and $\phi$, and the high-virtuality ``anomalous'' part either 
with towers of excited vector mesons or with perturbative $\q\qbar$
states. 
\item Each vector-meson class may in its turn have ``elastic'',
diffractive and non-diffractive topologies. 
\item The jet cross section in the non-diffractive component presumably 
is what drives the energy dependence of the total cross section, but 
contributions may additionally come e.g. from soft pomeron exchange. 
The perturbative jet cross section is divergent in the limit 
$\pT \to 0$, so some nonperturbatively motivated cut-off procedure is 
required. 
\item When the jet cross section is large, the possibility of multiple 
parton--parton interactions is non-negligible, and an eikonalization 
approach can be invoked to address the multiplicity distribution of
interactions and the mixing of event classes. The definition of the 
multiparton distributions of the p and $\gamma$ is here far from obvious. 
\item Each parton--parton interaction is associated with the possibility
of further radiation, to be calculated either in a matrix-elements 
approach or by invoking initial- and final-state parton showers.
Issues include matching of hard scattering with showers, coherence effects,
and cut-offs. 
\item The proton and a resolved photon contains a beam jet, still not
well understood. The remnant takes ``what is left'' after the hard
interaction(s) with associated radiation, including a ``primordial
$k_{\perp}$'' recoil. The latter is expected to be larger for the 
higher-mass anomalous states than for events associated with the 
lowest-lying vector mesons.
\item So far, only parton production has been considered. Confinement
implies that these hadronize to produce a set of primany hadrons that
can subsequently decay further.
\end{Itemize} 
The multitude and complexity of tasks reduces the scope for purely
analytical studies. On the other hand, the above subdivision may be 
seen as a suitable starting point for the construction of event
generators. Today programs such as {\small HERWIG} \cite{Herwig}, 
{\sc Phojet} \cite{Phojet} and {\sc Pythia} \cite{Pythia,gammap} are used
frequently to compare with data and to extract physics conclusions.
 
In view of the multitude of topics, only some of these will be covered
in the following.
 
\section{Hadronization}

These days the issue of hadronization is considered so standard
that it is not very much discussed. The recent measurements of
strangeness production at HERA have brought the issue back to
focus, as follows.

In our standard approach to multiparticle production, we often assume 
that hadronization is a universal process that can be factorized from 
a preceding perturbative stage. If so, the free parameters of a 
hadronization model can be determined once and for all, e.g. from 
e$^+$e$^-$ data at LEP, and thereafter applied to HERA events. Such a 
factorization is explicit in the Lund model \cite{Lund}, where all 
hadron production is caused by 
the stretching and breaking of strings. Since there is only one kind
of string, in principle it is only necessary to specify the parton 
content and string drawing topology (colour connectedness) of events to 
predict the structure of hadronic final states. In practice, some new
aspects of nonperturbative physics do appear in ep/$\gamma$p, such as 
parton distributions, multiple interactions and the treatment of beam  
remnants. An imperfect modelling either of perturbative or of
nonperturbative effects could show up in a mismatch in the total number
of hadrons produced, while the relative composition of different
hadron species should be rather stable.

It is here that the HERA data provide a surprise. ZEUS \cite{sZEUS} and 
H1 \cite{sH1} both observe a deficit of strangeness production. 
Interpreted in string fragmentation terms, the s/u ratio is shown to be
of the order of 0.2 rather than the 0.3 normally observed in e$^+$e$^-$ 
data (see e.g. \cite{LEP2QG} and references therein). The ZEUS study is 
for DIS, while the H1 one covers also photoproduction. The suppression
affects both K$^0_{\mrm{S}}$ and $\Lambda$ production.

In fact, the problem is not quite new, but has been observed before at
fixed-target energies, in neutrino and muon interactions \cite{sold}.
However, earlier data were partly contradictory \cite{sold}, and so
we tended to think of it as some specific low-energy
problems and not care so much. Obviously, this will not work now.
Although not necessarily related, one should also keep in mind two 
possible anomalies in flavour production at LEP: a ``strangeness deficit'' 
in the subclass of very two-jetlike events \cite{sDELPHI} and an 
$\eta$ production ``excess'' in the gluon jet in three-jet events 
\cite{sL3}.

Other existing models might solve the problem at HERA. 
In an approach such as {\sc Phojet}, the production of multiple
small chains with a non-negligible phase-space suppression in each 
can make a difference. In the {\small HERWIG} approach there is no
explicit jet universality: branchings $\g \to \q\qbar$ are
used to split the partonic final state into clusters that then
produce the hadrons according to phase space, so the cluster mass 
spectrum directly influences
the particle composition. Based on our current 
understanding, the effect would go in the wrong direction:
the cluster mass spectrum is universal for clusters produced by
the perturbative cascades, but these cascades are suppressed close
to the beam remnants, and this leads to larger remnant cluster masses 
with the possibility of enhanced strangeness and baryon 
production. Some additional source of soft-gluon emission close to the
beam remnants could revert this trend, however.

Within the Lund string model, it is interesting to speculate on a true
breakdown of jet universality. Here comes three examples of possible
physics mechanisms:
\begin{Itemize}
\item The ``quiet string scenario''. A conventional QCD cascade gives a
fractal structure \cite{Gosta}, i.e. the string is wrinked on all scales
(down to some unknown infrared cut-off).
The string tension $\kappa \approx 1$~GeV/fm is an effective parameter 
based on measurements at hadronic distance scales. The ``true'' string 
length, defined along all the wrinkes, is larger than the smoothened-out
normal length, and therefore the ``bare'' string tension is 
correspondingly smaller. If the amount of soft-gluon radiation is
smaller in ep than in e$^+$e$^-$, it would lead to a 
less wrinkled string and therefore a smaller effective string tension.
The string tension appears in the tunnelling mechanism of flavour 
production \cite{Lund}, and so a smaller $\kappa$ is directly reflected 
in a reduced strangeness production. A prediction in such a scenario is 
that baryon production should be even more suppressed than strangeness is.
\item Medium dependence. Unlike e$^+$e$^-$, the ep perturbative processes 
appear inside the ``hadronic bag'' of the proton, so why could not this
affect particle production? The counterargument would be that hadronization 
is a long-distance process, that only appears once the string is stretched
beyond the confines of the original proton, so somehow information would
have to survive a long distance. 
\item A separate kind of gluon string. This is allowed within a fairly
standard extension of the normal string model \cite{Montvay} and could
be combined with a string fragmentation model giving more glueballs
\cite{PetWal} and maybe also strangeness. So a larger amount of energetic
gluons in e$^+$e$^-$ than in ep could induce some of the desired 
difference.
\end{Itemize}

To be honest, neither of the above approaches appears particularly 
attractive, with the first the least contrived. 
Further experimental input therefore is eagerly awaited. 
First, the observations should be verified and extended
to more hadron species, especially (anti)baryons. Second, ratios such as 
K/all are preferrable, since then theory uncertainties in the total 
multiplicity divide out. Third, in DIS events a comparison of the
current and target hemispheres in the Breit frame would be revealing
--- any difference between the current hemisphere and e$^+$e$^-$ would
be very difficult to explain away. Fourth, in photoproduction a 
corresponding subdivision would be into high-$\pT$ jets and beam jets, 
again with the former more constrained by jet universality arguments. 
Fifth, can the strangeness deficit be related with any other property 
of events, so that a pattern emerges?

\section{Event Classes}

The photon can fluctuate into charged fermion--antifermion pairs.
Low-virtuality fluctuations may be associated with a sum over vector-meson
states while high-virtuality fluctuations are better described by
a continuous spectrum of states. A convenient ansatz for the photon 
wave function then is \cite{gammap}
\begin{equation}
|\gamma\rangle = c_{\mrm{bare}} |\gamma_{\mrm{bare}}\rangle +
\sum_{V = \rho^0, \omega, \phi, \Jpsi} c_V |V\rangle +
\sum_{\q = \u, \d, \s, \c, \b} c_{\q} |\q\qbar\rangle +
\sum_{\ell = \e, \mu, \tau} c_{\ell} |\ell^+\ell^-\rangle 
\label{gammawavefunction}
\end{equation} 
(neglecting the small contribution from $\Upsilon$). In general, the 
coefficients $c_i$ depend on the scale $\mu$ used to probe the photon.
Thus $c_{\ell}^2 \approx (\alphaem/2\pi)(2/3) \ln(\mu^2/m_{\ell}^2)$. 
Introducing a cut-off parameter $k_0$ to separate the low- and 
high-virtuality parts of the $\q\qbar$ fluctuations, one similarly 
obtains $c_{\q}^2 \approx (\alphaem/2\pi) 2e_{\q}^2 \ln(\mu^2/k_0^2)$.
The VMD part corresponds to the range of $\q\qbar$ fluctuations below
$k_0$ and is thus $\mu$-independent (assuming $\mu > k_0$). 
Finally, $c_{\mrm{bare}}$ is given by
unitarity: $c_{\mrm{bare}}^2 \equiv Z_3 = 1 - \sum c_V^2 -
\sum c_{\q}^2 - \sum c_{\ell}^2$. In practice, $c_{\mrm{bare}}$ is
always close to unity.

The leptonic component is not interesting for strong-interaction 
physics, but the other three can be associated with the direct, VMD 
and anomalous event classes. All three processes are of $O(\alphaem)$. 
However, in the direct contribution the photon structure function is 
of $O(1)$ and the hard scattering matrix elements of $O(\alphaem)$, 
while the opposite holds for the VMD and the anomalous processes. 

The separation is very convenient in a leading-order description, but
in higher-order contributions the various components appear mixed.
This certainly is a complication, but it should not be over-stressed.
We expect that the (numerically) dominant contributions will come from
event topologies that can still be classified as above. For instance,
in lowest order the direct process is characterized by the complete 
absence of a beam remnant in the photon direction, while some energy flow 
can always be expected in higher orders. Such a smearing is already 
included in generators by the addition of standard (coherent) 
parton-shower activity. To leading-log accuracy, the direct process 
then is characterized by a ladder diagram where the largest 
virtuality is found in the ladder adjacent to the photon, while resolved 
proceses have the largest virtuality somewhere in between the photon
and the proton, with decreasing virtualities on either side.
Alternatively, it is possible to imagine functional
separations, with some fraction $\epsilon$ of energy allowed within a
cone $\delta$ around the beam direction for direct processes, along
what is done experimentally.  

Whichever approach is adopted, it is important that physics should impose 
a smooth joining between the event classes. Any classification is a
matter of convenience. However, at our current level of understanding, 
phenomenological studies can be made more realistic if they are based 
on a pragmatic division of $\gamma$p (and $\gamma\gamma$ \cite{gaga}) 
events into separate subclasses.

\section{The Total Cross Section}

There are two common approaches to the issue of the total cross section
in $\gamma$p (as well as pp and $\gamma\gamma$) collisions. One is 
the Regge-theory ansatz, where $\sigma_{\mrm{tot}}$ is given as the sum 
of two terms, the pomeron one ($\propto s^{\epsilon}$, 
$\epsilon \approx 0.08$) and the reggeon one 
($\propto s^{-\eta}$, $\eta \approx 0.45$) \cite{DoLa}. This ansatz 
gives a very handy parametrization of cross sections, that seems to be 
in good agreement with data. However, it does not necessarily lead to 
any understanding of the underlying physics.

More appealing is the second main approach, where the rise of the total 
cross section at large energies is related to the increase of the jet
cross section. In its simplest variant, one would write
$\sigma_{\mrm{tot}}(s) = \sigma_{\mrm{soft}}(s) + 
\sigma_{\mrm{jet}}(s, \pTmin)$ \cite{drees}. The $\sigma_{\mrm{jet}}$
term is obtained by integrating the perturbative $2\to 2$ 
hard-scattering cross section in the region 
$p_{\perp} > \pTmin$. Uncertainties come from 
the choice of $\pTmin$ scale, from parton distributions,
from higher-order corrections to the lowest-order matrix elements, 
from the choice of a $\sigma_{\mrm{soft}}(s)$, and so on. Furthermore, 
if one attempts to limit the arbitrariness by keeping 
$\pTmin$ independent of $s$, the approach breaks down at
large energies, where the jet cross section is known to increase faster
than the total one. 

We understand that this is linked to the emergence
of events with several parton--parton interactions above the
$\pTmin$ scale. For instance, an event with two interactions should
count twice against the hard-scattering cross section, but only once 
against the total one. The eikonalization approach is a convenient
way of accounting for an arbitrary number of interactions. Normally 
the direct processes are assumed unaffected, i.e. only the ones with
a resolved photon are eikonalized. In addition to the input already 
mentioned, one here needs to specify the probability for a photon to 
turn into a hadron \cite{collins}, the impact parameter
dependence of the eikonal (obtained as a convolution of the matter
densities of the two incoming particles), the r\^ole of elastic and
diffractive topologies, and so on. Sub-variants are possible, such as 
leaving $\sigma_{\mrm{soft}}$ out of the eikonalization machinery
\cite{forshaw}.

In a further level of sophistication, the probability for a photon to
interact like a hadron can be replaced by a sum over discrete
vector-meson states plus an integral over a continuum of perturbative
$\q\qbar$ states (the anomalous component) \cite{sarcevic}. Each state 
is now to be eikonalized separately, and each with its own set of free
parameters: soft cross sections, matter densities, and so on.
The only area where the freedom is reduced by this choice is for 
parton distributions, where the VMD ones in principle are measurable
(though in practice not, so one uses e.g. the $\pi$ ones) and the
anomalous ones are calculable. 

Unless the energy-dependence of $\sigma_{\mrm{soft}}$ is fine-tuned,
it is difficult to obtain a turnover from a falling to an increasing
$\sigma_{\mrm{tot}}(s)$ at as low energies ($\sqrt{s} \simeq 10$~GeV)
as observed experimentally, simply because the jet rate above
some reasonable $\pTmin \gtrsim 1$~GeV only picks up at larger energies.
It appears plausible that soft, nonperturbative multiple interactions 
in fact drive the change of $\sigma_{\mrm{tot}}$ at low energies. 
One attractive framework for putting it all together is DTU 
(dual topological unitarization), where both the soft and hard
interactions, the triple- and loop-pomeron graphs responsible for 
diffractive topologies, and higher-order pomeron graphs are all
put together \cite{DTU}. New parameters include several pomeron and
reggeon couplings.
 
In the end, even this complex machinery is hardly more successful than 
the simple Regge-theory-based one we started out with. In fact, if the 
only criterion is predictive power for the total cross section at higher 
energies, it could be argued that the simple pomeron-type ansatz is the 
best bet. Somewhat surprisingly, the experience outlined above teaches us 
that there is a tradeoff between sophistication and predictive power: the 
more advanced we try to be, the more free parameters we have to play
with, and the less constrained we are about what will happen at energies
not yet explored.

So when we still persevere to build ever more detailed models for the 
total cross section, it is because the ultimate goal is to reach an 
understanding of the nature of the photon and its interactions.
If we have reasons to believe that the photon has a complex nature, 
then we should not expect to get away with simple recipes for 
everything. A sophisticated approach also provides a blueprint for 
how to model or predict a number of exclusive event properties.
Testability therefore comes not only from the total cross section.

One test is provided by partial cross sections to ``elastic'' and 
diffractive topologies. The elastic process 
$\gamma\p \to \rho^0\p$ turns out to be very well predicted, both
the cross section and the $t$ slope. This certainly is a major
success for the VMD approach: the picture with 
$\gamma \leftrightarrow V$ fluctuations is now shown to hold 
independently of energy. Discrimination between models is obtained
by the H1 study of diffractive cross sections \cite{H1dif}. Here
the DTU approach \cite{DTU} does best among the ones studied;
specifically, it correctly predicts that the ``photon-diffractive''
process $\gamma\p \to X\p$ has a much larger cross section than
the ``proton-diffractive'' one $\gamma\p \to VX$. Many further tests
can be expected in the years to come.

\section{Multiple Interactions}

The importance of multiple parton--parton interactions is more 
discussed and established
among HERA physicists than it is among $\p\pbar$ collider people,
in spite of the larger energies available to the latter. This is an
interesting paradox, which can only be understood if one considers 
the ``historical prejudices'' of the two communities. DESY has a 
background in e$^+$e$^-$ physics, where the combination of a perturbative
(shower) picture and a universal hadronization stage is firmly 
established. When such a picture applied to photoproduction events
gives too little activity at small transverse momenta, both in 
``minimum-bias'' events and in the ``underlying event'' of jets,
it is therefore interpreted as a sign of additional 
(semi)perturbative activity with its associated universal 
hadronization, in line with predictions \cite{gammap}.

Experiments at hadron colliders, on the other hand, have a tradition
stretching back to before the days of QCD-based models for
multihadron production. Therefore another philosophy has 
developed in that field: high-$\pT$ jets are considered as
standard QCD objects but the low-$\pT$ activity is described 
in terms detached from any jet universality constraints. A common 
attitude is that ``soft hadronic physics
is so dirty that you cannot predict anything; let us therefore
simply parametrize whatever we observe''. The  
litmus test of multiple interactions, namely the observation of an
excess of two jet pairs in the same event, with the pairs identified
e.g. by each having vanishing net transverse momentum, is
very difficult experimentally. Therefore studies have not been conclusive, 
though the picture with multiple interactions is favoured
\cite{multintdata}. And, strictly speaking, the observation of multiple
interactions at moderately large $\pT$ does not tell anything about their
possible r\^ole in the soft region.

Attempts to produce support for multiple interactions based on 
jet universality arguments \cite{TSMZ} have not caught on in the
$\p\pbar$ community, though the ``evidence'' is reasonably compelling 
(in the eyes of the believers). To give a few examples:
\begin{Itemize}
\item The charged multiplicity and the transverse energy is increasing
with energy much faster than the $\ln s$ that could be expected 
from the increasing rapidity range.
\item The multiplicity distribution is much broader than the roughly
Poissonian shape that is predicted from a single (or double) string.
A reasonable account of the experimental ``negative Binomial'' 
distribution, with a relative width 
$\sigma(n_{\mrm{ch}}) / \langle n_{\mrm{ch}} \rangle$ that increases
with energy, can be obtained by adding the further element of 
randomness caused by a variable number of semihard interactions
in events.  
\item The data also contains strong forward--backward multiplicity
correlations: if one hemisphere of an event has a large multiplicity
then, normally, so does the other. A varying number of strings,
frequently stretched over both hemispheres, easily explains this
phenomenon.
\end{Itemize}  
 
In spite of the lower energy, HERA has a chance to provide many further
interesting tests of multiple interactions, and put the whole game on
much firmer footing. The main reason is the variability offered by the
photon probe:
\begin{Enumerate}
\item Multiple interactions are expected to vanish gradually as the photon
virtuality $Q^2$ is increased. This may be seen as a consequence of the 
reduced number of (resolved) partons in a higher-virtuality photon. 
\item The direct events are not expected to contain multiple interactions.
This can be observed as a decrease of multiplicity and $E_{\perp}$
for events with larger $x_{\gamma}$.
\item Within the resolved class, the anomalous events are expected to
have less multiple interactions than the VMD ones. This may be understood
by considering a $\gamma \to \q\qbar$ branching at a transverse momentum
$k_{\perp}$, where the latter quantity in principle is measurable from
the $\pT$ of the remnant jet. The $k_{\perp}$ sets a virtuality
scale, like $Q$ in the first point above, with reduced evolution
range and therefore fewer partons in a photon branching at a larger 
$k_{\perp}$. Additionally, the $\pTmin$ cut-off of (semi)hard interactions
can be expected to increase with $k_{\perp}$, thus further reducing
multiple interactions.
\item Points 2 and 3 above come together in the variation of the
multiple-interaction rate as a function of the $\pT$ of an observed
jet. It is here well-known that a larger $p_{\perp\mrm{jet}}$ biases
the event sample towards direct and anomalous events, and hence
should give fewer multiple interactions. This ``anti-pedestal'' effect 
should take over at larger $p_{\perp\mrm{jet}}$, whereas the smaller 
$p_{\perp\mrm{jet}}$ events should show the conventional pedestal
effect \cite{UA1ped} presumably caused by an impact-parameter variation
\cite{TSMZ}.
\item Multiple interactions could offer a chance to probe ``hot spots''
in the proton. A virtual photon with virtuality $Q \gtrsim m_{\rho}$
probes a region of size $\sim 1/Q$, and a (real) anomalous photon with
a branching $k_{\perp} \gtrsim m_{\rho}$ probes a region of size
$\sim 1/k_{\perp}$, so by increasing $Q$ or $k_{\perp}$ a smaller region
of the proton is probed. As discussed above, the multiple interaction
rate should go down in either case, but the question is whether it does 
so uniformly.
If the proton contains hot spots, with several nearby partons, a photon
probe hitting such a spot will still have a non-negligible chance of
multiple interactions. In terms of an inclusive distribution of the 
charged multiplicity or summed $E_{\perp}$, the average value 
should be independent of the existence of hot spots, but the 
event-by-event fluctuations
around this average would go up with hot spots present. Results on this
topic would tie in with the small-$x$ behaviour of parton distributions
and saturation effects.   
\end{Enumerate}

\section{Summary} 

After a few years of exciting HERA results, it is clear that existing
models do a reasonable job of explaining the data. In this sense, we
do have a zeroth approximation to work with. This is always useful as 
a guide to help us classify and understand phenomena, but it should
not straight-jacket our thinking. Moreover, agreement between data and 
models is far from perfect, so there is no reason for complacency.
There are several areas where more 
work is needed to see whether we actually have all the necessary
ingredients at hand. It is in no sense excluded that we need to
develop new ways of thinking. Among the many issues one may mention:
\begin{Itemize}
\item jet universality and systematic comparisons with e$^+$e$^-$,
$\p\pbar$, DIS $\gamma\p$ and $\gamma\gamma$ events;
\item the change in event properties (such as rapidity gaps) when
moving from a real to a virtual photon;
\item multiple interactions and hot spots;
\item the smooth joining of event classes;
\item the character of beam jets; and
\item the mass spectrum of diffractive states.
\end{Itemize}   
In view of this, it is important to remember that we are only at the
beginning of the story.

\vspace{\baselineskip}\noindent
\textbf{Acknowledgments:} The organizers are thanked for a very  
stimulating workshop, and Gerhard Schuler for an enjoyable 
collaboration.

\end{document}